\documentclass[11pt,notoc,hyper]{article}
\usepackage{jcappub}

\usepackage{bm,amsmath,amssymb}
\usepackage{dcolumn}
\usepackage{epsfig,multicol,float} 

\def\simgt{\lower.5ex\hbox{$\; \buildrel > \over \sim \;$}}
\def\simlt{\lower.5ex\hbox{$\; \buildrel < \over \sim \;$}}

\def\({\left(}
\def\){\right)}

\def\integ0a{\int_0^a}
\def\[{\left[}
\def\]{\right]}

\newcommand{\ang}{\hat\nabla}
\newcommand{\up}[1]{{\rm #1}}
\newcommand{\bdv}[1]{{\bf #1}}

\newcommand{\beeq}{\begin{equation}} 
\newcommand{\eneq}{\end{equation}}
\newcommand{\bear}{\begin{eqnarray}}
\newcommand{\enar}{\end{eqnarray}}

\newcommand{\vL}{{\bf l}}
\newcommand{\Vang}{\bdv{\hat n}}

\newcommand{\lenX}{\tilde X}

\newcommand{\CT}{C^T}
\newcommand{\CE}{C^E}

\newcommand{\CC}{C^C}
\newcommand{\lenCX}{\tilde C^X}

\newcommand{\Xobs}{\tilde X^{\up{obs}}}
\newcommand{\CXobs}{\tilde C^{X,\up{obs}}}
\newcommand{\Yobs}{\tilde Y^{\up{obs}}}
\newcommand{\CYobs}{\tilde C^{Y,\up{obs}}}
\newcommand{\GXY}{\bdv{G}_{XY}}
\newcommand{\WY}{W_Y}
\newcommand{\khat}{\hat\kappa}
\newcommand{\Ckap}{C^{\kappa}}
\newcommand{\Nkap}{N^{\kappa,XY}}

\newcommand{\muK}{\mu\up{K}}

\newcommand{\spix}{\sigma_{\up{pix}}}
\newcommand{\opix}{\Omega_{\up{pix}}}

\newcommand{\sbeam}{\sigma_b}
\newcommand{\tfwhm}{\theta_{\up{FWHM}}}

\newcommand{\bn}{\hat{\bf n}}
\newcommand{\bl}{{\bf l}}
\newcommand{\bll}{{\bf L}}

\newcommand{\intlp}[1]{\int {d^2 l_{#1}' \over (2\pi)^2}}

\def\l#1{\ell_#1}

\newcommand{\reffig}[1]{Figure~\ref{#1}}

\newcommand{\eqn}[1]{Eq.~(\ref{#1})}

\def\be{\begin{equation}}
\def\ee{\end{equation}}
\def\bea{\begin{eqnarray}}
\def\eea{\end{eqnarray}}

\def\simlt{\lower.5ex\hbox{$\; \buildrel < \over \sim \;$}}
\def\simgt{\lower.5ex\hbox{$\; \buildrel > \over \sim \;$}}
\def\simgtalt{\lower.5ex\hbox{$\buildrel > \over \sim \;$}}
\def\l#1{\left#1}
\def\r#1{\right#1}

\def\bd#1{\bm{#1}}

\def\figdir#1{figures/#1}

\title{CMB lensing reconstruction in the presence of diffuse polarized
foregrounds}

\author[a]{Y.~Fantaye,}
\author[a,b]{C.~Baccigalupi,}
\author[a]{S.~M.~Leach}
\author[c]{and A.~P.~S.~Yadav}

\affiliation[a]{SISSA, Astrophysics Sector, via Bonomea 265, Trieste 34136, Italy} 
\affiliation[b]{INFN, Sezione di Trieste, Via Valerio 2, I-34151 Trieste, Italy} 
\affiliation[c]{Center for Astrophysics and Space Sciences, Department of Physics, University of California, San Diego, 9500 Gilman Drive, La Jolla, CA, 92093-0424}

\emailAdd{fantaye@sissa.it}
\emailAdd{ bacci@sissa.it} 
\emailAdd{leach@sissa.it} 
\emailAdd{ayadav@physics.ucsd.edu}

\abstract{The measurement and characterization of the lensing of the
  cosmic microwave background (CMB) is key goal of the current and
  next generation of CMB experiments. We perform a case study of a
  three-channel balloon-borne CMB experiment observing the sky at
  ($l$,$b$)=($250^\circ$,$-38^\circ$) and attaining a sensitivity of
  5.25 $\muK-$arcmin with $8'$ angular resolution at 150 GHz, in order
  to assess whether the effect of polarized Galactic dust is expected
  to be a significant contaminant to the lensing signal reconstructed
  using the $EB$ quadratic estimator. We find that for our assumed
  dust model, polarization fractions of about as low as a few percent
  may lead to a significant dust bias to the lensing convergence power
  spectrum. We investigated a parametric component separation
  method, proposed by Stompor et al. (2009), as well as a template
  cleaning method, for mitigating the effect of this dust bias. The
  template-based method recovers unbiased convergence power spectrum
  in all polarization fraction cases we considered, while for the
  component separation technique we find a dust contrast regime in
  which the accuracy of the profile likelihood spectral index estimate
  breaks down, and in which external information on the dust frequency
  scaling is needed. We propose a criterion for putting a requirement
  on the accuracy with which the dust spectral index must be estimated
  or constrained, and demonstrate that if this requirement is met,
  then the dust bias can be removed.}

\begin{document}

\maketitle

\section{Introduction}
The measurement and characterizaton of the weak lensing of the cosmic
microwave background (CMB) by the large-scale structure distribution
is a promising and active field of research in observational cosmology
(for a review of the physics of CMB weak lensing, see
\cite{Lewis:2006}). Measurements of this signal can break
fundamental degeneracies that afflict the cosmological interpretation
of measurements of the CMB power spectrum~\citep{Stompor:1999mnrs} as
well as help to improve the constraints on the cosmological
parameters~\citep{Perotto_FutureCMB_06,2012arXiv1205.0474B}. As a
result, a number of ongoing and planned experiments are targeting the
weak lensing signal as one of their primary science goals.

The CMB weak lensing signal was first detected
by~\cite{2007PhRvD..76d3510S} who cross correlated data from the
Wilkinson Microwave Anisotropy Probe (WMAP) satellite with a tracer of
large-scale structure in the form of the NRAO VLA sky survey. Since
then observational progress has been rapid: the Atacama Cosmology
Telescope (ACT) collaboration made the first detection of weak lensing
signal using CMB data alone~\citep{das2011} and the South Pole
Telescope (SPT) collaboration have followed with a detection at higher
significance~\citep{Engelen:2012spt} as well as detecting the
correlation of the weak lensing `convergence' and large-scale
structure tracers from the Wide-field Infrared Survey Explorer and
Spitzer/IRAC~\citep{2012arXiv1203.4808B}. First applications of the
weak lensing signal measurements have been to provide corroborating
evidence for the cosmological constant from CMB data
alone~\citep{Sherwin:2011gv} and to improve constraints on the dark
energy equation of state~\citep{Engelen:2012spt}.  In the future,
improved cosmological constraints are expected from the full SPT and
ACT surveys, and especially from \emph{Planck} which is poised to
significantly advance lensing
studies~\citep{Perotto:2009tv,2011PhRvD..83d3005H}.

The most widely adopted technique for extracting the lensing signal, also
implemented in this investigation, is the `quadratic estimator'
suggested by~\cite{HuOkamoto:2002} which is a near optimal framework
for reconstructing the lensing field using a quadratic combination of
an appropriately filtered CMB temperature and/or polarization maps and its
gradient. The subsequent estimation of the power spectrum of the
lensing field is met by a variety of real world complications. These
include intrinsic biases of the quadratic estimator which must be
modeled and
subtracted~\citep{2003PhRvD..67l3507K,2011PhRvD..83d3005H}, and any
effect that breaks the statistical isotropy of data including
anisotropic noise~\citep{2009MNRAS.400.2169H} and the effect of
instrumental
systematics~\citep{2009PhRvD..79f3008M,2009PhRvD..79l3002S}.

Astrophysical foregrounds represent another source of contamination to
the lensing signal for which there have been several simulation
studies assessing their possible impact and suggesting mitigation
strategies. \citep{Amblard:2004ih} performed simulations of the
kinetic Sunyaev-Zeldovich (SZ) effect and pointed out that this
non-Gaussian effect correlated with the lensing signal, with a
spectrum indistinguishable from CMB anisotropies, could potentially
bias the lensing reconstruction. They proposed masking CMB
maps around the location of clusters detected via the thermal SZ
effect as a potential strategy for mitigating kSZ biases, a procedure
subsequently adopted by the SPT
team~\cite{Engelen:2012spt}. \cite{Smith:2009cmbpol} made calculation
of the lensing biases induced by polarized extragalactic radio point
sources, concluding that the effect of these source could be mitigated
via source detection/masking, and that these sources ought not pose a
significant challenge to a future CMB polarization
satellite. \cite{Perotto:2009tv} studied the impact of diffuse and
extragalactic foregrounds on a \emph{Planck}-like simulations and
proposed a multi-frequency masking and `in-painting' technique for
measuring the lensing signal.

To date there has been no specific study on the possible impact of
Galactic polarized dust emission on the detection of the lensing
signal. We believe that it is important to make such an assessment
given the fact that several ongoing ground-based and balloon-borne CMB
polarization experiments including ACTPol~\citep{2010SPIE.7741E..51N},
SPTPol~\citep{2009AIPC.1185..511M}, EBEX~\citep{2010SPIE.7741E..37R}
and POLARBEAR~\citep{polarbear} are targeting the lensing signal in the
near future, using measurements of CMB polarization in the frequency
range 90--410 GHz. In particular we will focus on performing a case
study of an EBEX-like experimental configuration which, owing to
observing constraints on balloon-borne telescopes flying from
Antarctica, will most likely survey a region of sky where Galactic
foregrounds in the form of polarized dust emission may be significant
compared to the lensing signal. We aim to develop methods for
numerically investigating the extent to which this experiment can
expect to be challenged by diffuse polarized dust emission, and to
investigate possible dust mitigation strategies.

This study is set out as follows: Section 2 describes the data
analysis techniques that we have implemented; Section 3 describes the
CMB and dust polarization simulations that we have performed; Section
4 describes our findings and results, and Section 5 contains our
conclusions.

\section{Data analysis techniques}
In this section we review the formalism of the two analysis techniques
that we have implemented and investigated in our lensing estimation
pipeline. These are the quadratic estimator~\cite{HuOkamoto:2002} for
estimating the CMB lensing signal, and the multi-frequency analysis
technique of~\cite{Stompor_etal_2009} for estimating the foreground
spectral index and performing separation of the CMB component.

\subsection{CMB lensing and the quadratic estimator} 
CMB photons from the last scattering surface are deflected according
to
\bear
\label{eq:lensing}
\tilde T(\bn)&=&T\left[\bn+\ang\phi(\bn)\right],\\
(\tilde{Q}+i\tilde{U})(\bn)&=&(Q\pm iU)\left[\bn+\ang\phi(\bn)\right],
\nonumber
\enar
where $\phi(\hat{\bf n})$ is the projected potential, and $\bn$ is the
direction on the sky~\citep{Zaldarriaga:1998te}. We will adopt the
notation convention that lensed quantities are denoted with a tilde
and unlensed ones without. In the Born approximation, in which the
deflection angles are assumed to be small enough to carry out the
projection kernel integration along the line of sight, $\phi$ is
related to the three-dimensional gravitational potential, $\psi$, by
\begin{eqnarray}
\phi(\bn) &=& -2\int_0^{D_\star}dD~{D_\star-D\over D D_\star}~\psi~(D\bn,D),
		\nonumber\\
 &=& 
		\int { d^2 L \over (2\pi)^2}
		\phi(\bll) e^{i \bll \cdot \bn} , 
\end{eqnarray}
where $D$ and $D_\star$ are the comoving angular diameter distances to
the lens and the CMB last scattering surface respectively; $\phi$ is
related to the `convergence' $\kappa$ as $\ang^2\phi=-2\kappa$
\citep{Zaldarriaga:1998te}.

To linear order in~$\phi$, the changes in the Fourier moments of the
temperature and polarization fields due to lensing are~\citep{Hu:2000ee}
\begin{eqnarray}
\delta \tilde T(\bl) &=& \intlp{} T(\bl') W(\bl',\bll),\\
\label{eqn:lensedl}
\delta \tilde E(\bl)      &=& \intlp{} 
\Big[ E(\bl') \cos 2\varphi_{\bl'\bl}
     - B(\bl') \sin 2\varphi_{\bl'\bl} \Big]
W(\bl',\bll),\nonumber\\
\delta \tilde B(\bl)      &=& \intlp{} 
\Big[ B(\bl') \cos 2\varphi_{\bl'\bl}
     + E(\bl') \sin 2\varphi_{\bl'\bl}\Big]
W(\bl',\bll),\nonumber
\end{eqnarray}	
where the azimuthal angle difference $\varphi_{\bl'\bl} \equiv \varphi_{\bl'} - \varphi_{\bl}$,
$\bll = \bl - \bl'$, and
\begin{eqnarray}
W(\bl,\bll) = -[{\bl \cdot \bll}]\phi(\bll).
\end{eqnarray}
From these equations we note that the effect of lensing is to couple
the gradient of the primordial CMB $\bl'$ modes to that of the
observed $\bl$ modes. Furthermore, starting from zero primordial
$B$-modes, $ B(\bl')=0$, lensing generates $B$-mode anisotropies
in the observed map.

\paragraph{Quadratic estimators for the convergence: }

We will assume a CMB map with homogeneous white noise and a Gaussian
beam smoothing. The power spectrum of the detector noise
is~\citep{Knox_95}
\begin{eqnarray}
C_{l}^{N,X}=\spix^2~\opix,
\label{eq:nois}
\end{eqnarray}
where $\spix$ is the RMS noise per pixel and $\opix$ is the solid
angle subtended by each pixel. The observed CMB temperature and
polarization fields, $X\in[T,E,B]$, and their power spectra,
$\tilde C_{\ell}^X$, are \bear
\label{eq:tobs}
\Xobs_\vL&=&\lenX_\vL~e^{-{1\over2}l^2\sbeam^2}+N^X_\vL,\\
\CXobs_l&=&\lenCX_le^{-l^2\sbeam^2}+C^{N,X}_l, \nonumber
\enar
where $N^X_\vL$ is the Fourier mode of the detector noise, and $\sbeam$
relates to the full-width half-maximum (FWHM) of the telescope beam,
$\theta_{\rm FWHM}$, via $\tfwhm=\sbeam\sqrt{8\ln2}$.

To extract the lensing information in the observed CMB map we will use
the quadratic estimator
formalism~\citep{Hu:2001fa,Hu:2001tn,HuOkamoto:2002} in the context of
the convergence estimators~\citep{Hu:2007njp, 2010PhRvD..81l3006Y}.
These estimators are uniquely determined by the
requirement that each estimator be unbiased
$\langle\khat^{XY}(\Vang)\rangle=\kappa(\Vang)$ over an ensemble
average of the CMB temperature and polarization fields $X$~and~$Y$,
and the variance of the estimator be minimal, 
\begin{eqnarray}
\langle\khat^{XY}_\vL\khat^{*XY}_{\vL'}\rangle=(2\pi)^2~\delta^D(\bdv{l-l'})
(\Ckap_l+\Nkap_l).
\label{eq:vari}
\end{eqnarray}
In real space the convergence estimators are given by
\cite{2010PhRvD..81l3006Y}
\bear
\label{eq:filter}
&&\GXY(\Vang)=\int\!\!\!{d^2\vL\over(2\pi)^2}~i\vL\Xobs_\vL
{C_l^{XY}\over\CXobs_l}
\bigg\{\begin{array}{c}e^{2i\varphi_{\vL}}\\e^{2i\varphi_{\vL}}\end{array}\bigg\}
e^{-{1\over2}l^2\sbeam^2+i\vL\cdot\Vang},\\
&&\WY(\Vang)=\int\!\!\!{d^2\vL\over(2\pi)^2}~{\Yobs_\vL\over\CYobs_l}
\bigg\{\begin{array}{c}e^{2i\varphi_{\vL}}\\ie^{2i\varphi_{\vL}}\end{array}\bigg\}
e^{-{1\over2}l^2\sbeam^2+i\vL\cdot\Vang},\label{eq:filter2}
\enar
where $\varphi_\vL$ is
the azimuthal angle of the wavevector~$\vL$; the two phase factors in
braces are applied when $Y=E,B$ respectively, and is unity when $Y=T$.  Also
$C_l^{XY}=C_l^{XE}$ for $Y=B$. The construction of these fields
incorporate the deconvolution of the beam from the map, hence the beam
factors $e^{-{1\over2}l^2\sbeam^2}$ appearing on both fields.
 
Given the two filtered fields in \eqn{eq:filter} and \eqn{eq:filter2}, 
the convergence estimators are then given by
\begin{eqnarray}
\khat^{XY}_\vL=-{A^{XY}_l\over2}~i\vL\cdot\int d^2\Vang~
\up{Re}\left[\GXY(\Vang)\WY^*(\Vang)\right]
~e^{-i\vL\cdot\Vang}.
\label{eq:GW}
\end{eqnarray}
The normalization coefficients, $A^{XY}_l$, are related to the noise power
spectrum, $\Nkap_l$, of the estimators $\khat^{XY}(\Vang)$ by
$\Nkap_l=l^2A^{XY}_l/4$, and are calculated as
\bear
\label{eq:nps}
{1\over A^{XY}_l}&=&{1\over
l^2}\int{d^2\vL_1\over(2\pi)^2}{(\vL\cdot\vL_1)~
C^{XY}_{l_1}f^{XY}_{\vL_1\vL_2}\over\CXobs_{l_1}~\CYobs_{l_2}} \\
&\times&
\bigg\{\begin{array}{c}\cos2\Delta\varphi\\\sin2\Delta\varphi\end{array}\bigg\}
~e^{-l_1^2\sbeam^2}~e^{-l_2^2\sbeam^2},\nonumber \enar with
$\vL=\vL_1+\vL_2$, $\Delta\varphi=\varphi_{\vL_1}-\varphi_{\vL_2}$,
and $\langle X_{\vL_1}Y_{\vL_2}\rangle=f_{\vL_1\vL_2}^{XY}~\phi_\vL$~,
where~\cite{Hu:2001tn}
\bear
\label{eq:fl}
f^{TT}_{\vL_1,\vL_2}&=&(\vL\cdot\vL_1)~\CT_{l_1}+(\vL\cdot\vL_2)~\CT_{l_2},
\\ f^{TE}_{\vL_1,\vL_2}&=&(\vL\cdot\vL_1)~\CC_{l_1}\cos2\Delta\varphi
+(\vL\cdot\vL_2)~\CC_{l_2}, \nonumber \\
f^{TB}_{\vL_1,\vL_2}&=&(\vL\cdot\vL_1)~\CC_{l_1}\sin
2\Delta\varphi,\nonumber \\
f^{EE}_{\vL_1,\vL_2}&=&\left[(\vL\cdot\vL_1)~\CE_{l_1}+(\vL\cdot\vL_2)~\CE_{l_2}
\right]\cos2\Delta\varphi,\nonumber \\
f^{EB}_{\vL_1,\vL_2}&=&(\vL\cdot\vL_1)~\CE_{l_1}\sin2\Delta\varphi.\nonumber
\enar
Our code for estimating the convergence using the quadratic
estimator formalism is a direct implementation of the above
equations, \eqn{eq:nois}--(\ref{eq:fl}).

\subsection{Foreground spectral index estimation and cleaning}
\label{sec:fgclean}

To perform foreground spectral index estimation and cleaning, we use
an implementation of the parametric component separation algorithm
proposed by~\citep{Stompor_etal_2009} and tested in the context of
forecasts for inflationary $B$-mode detection
in~\citep{Stivoli_etal_2010,2011JCAP...08..001F}.  In this framework,
the multi-frequency data vector $d_p$ is modeled at each pixel $p$
of the map as
\begin{equation}
d_p = A_p s_p + n_p,\label{eqn:datamodel}
\end{equation}
where $A_p\equiv A_p(\beta)$ is an $N_{\rm freq}\times N_{\rm comp}$
`mixing matrix' with $N_{\rm spec}$ free parameters, $\beta$, to be
estimated, $s_p$ is a vector of $N_{\rm comp}$ component amplitudes
also to be estimated, and $n_p$ is the noise. A likelihood for the
data is given by
\begin{equation}
-2 \ln \mathcal{L}(s,\beta) = {\rm CONST} +
(\bd{d}-\bd{As})^t \bd{N}^{-1}(\bd{d}-\bd{As}),
\end{equation}
where $\bd{N}^{-1}$ is the noise covariance of the map. A key result
of~\citep{Stompor_etal_2009} is that under the assumption that the
spectral index is the same for each pixel, the maximum-likelihood
values of the components $\bd{s}$ can be found by first locating the values
of the spectral parameters $\beta$ that  maximise the value of the
`profile likelihood'
\begin{equation}
-2\ln {\cal L}_{spec}\l(\bd{\beta}\r) =  \hbox{{\sc const}}
\label{eqn:profile}
-\l( \bd{A}^t \bd{N}^{-1}\bd{d}\r)^t\l(\bd{A}^t\bd{N}^{-1}\bd{A}\r)^{-1} \l( \bd{A}^t\bd{N}^{-1}\bd{d}\r),
\end{equation}
which is an expression independent of $\bd{s}$.  Once the maximum
likelihood spectral parameters, $\hat\beta$, have been determined,
their values are substituted into the generalized least squares
solution of Eq.~(\ref{eqn:datamodel}), given by
 \begin{equation}
\hspace{1cm} \bd{s} = \l(\bd{A}^t\bd{N}^{-1}\bd{A}\r)^{-1}\bd{A}^t\bd{N}^{-1}\bd{d},
\label{eqn:gls}
\end{equation}
\begin{equation}
\bd{N_s} \equiv \l(\bd{A}^t\bd{N}^{-1}\bd{A}\r)^{-1},
\label{eqn:noisecorr}
\end{equation}
to obtain the estimated component  amplitudes, $\bd{\hat s}$, and their
noise covariance $\bd{N_{\hat s}}$, pixel by pixel.


Insight into the component separation error can be gained using the
$\bd{Z}$-matrix formalism derived in~\cite{Stivoli_etal_2010} where,
\begin{eqnarray}
\bd{Z}(\bd{\hat\beta}) =  \left(\bd{A}^t(\bd{\hat\beta})\bd{N}^{-1}
\bd{A}(\bd{\hat\beta})\right)^{-1}\bd{A}^t(\bd{\hat\beta})\bd{N}^{-1}
\bd{A}\left(\bd{\beta_0}\right).
\label{eqn:zmat}
\end{eqnarray}
In our case, this will be a $2 \times 2$ matrix corresponding to the
two components, CMB and dust. In the limit of $\hat\beta=\beta_0$ then
$\bd{Z}$ is the identity matrix. For the case where the dust spectral index
is mis-estimated, then the off-diagonal terms of $\bd{Z}$ quantify the
fraction of the original component that remains unsubtracted, since
\begin{equation}\label{eqn:zsum}
  \bd{\hat s_i} = \sum_j^{n_{\rm comp}}{\bd{Z_{ij}(\hat \beta)s_j(\beta_0)}},
\end{equation}
where $\bd{\hat s}$ are the estimated components.

\section{CMB and polarized dust simulations}

\begin{figure}
\centering
\includegraphics[width=3in]{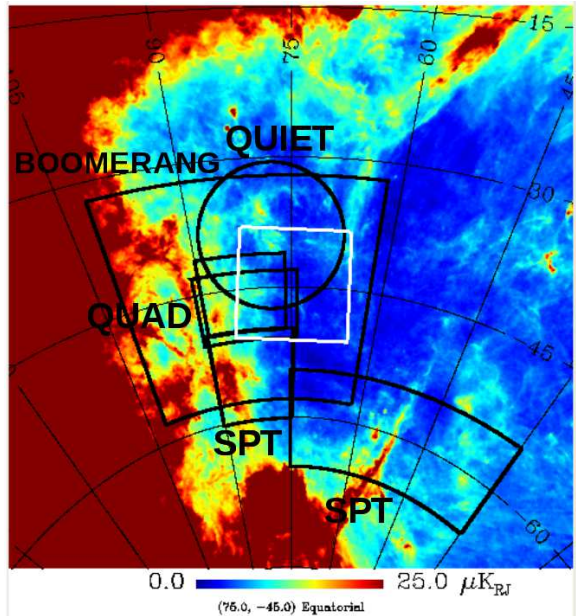} 
\caption{Dust intensity at 150 GHz (thresholded to 25 $\mu$K$_{\rm
RJ}$). Black outlines show the borders of previous CMB intensity and
polarization surveys, while the white outline shows the $13^\circ
\times 13^\circ$ patch that we consider in this study.}
\label{fig:skypatch}
\end{figure}

We simulate CMB polarization and diffuse polarized dust emission on a
$13^\circ \times 13^\circ$ patch of sky located at (RA, Dec) =
($75^\circ$,$-44.5^\circ$), corresponding to
($l$,$b$)=($250^\circ$,$-38^\circ$) in Galactic coordinates. This area
of sky is accessible from observing sites both in Antarctica and
Chile, has been surveyed by several past ground-based and
balloon-borne CMB experiments including Boomerang~\citep{Boomrang},
QUAD~\cite{quad_2009}, ACBAR~\citep{acbar},
QUIET~\cite{2011ApJ...741..111Q} and SPT~\citep{2011ApJ...743...28K},
as shown in Figure~\ref{fig:skypatch}. It is close to the area that
will be observed by the balloon-borne CMB experiment EBEX, owing to
the fact that is aligned with the anti-sun direction in mid December,
soon after the start of the Antarctic long duration balloon flight
launch window. Given that the anti-sun direction moves by 1 degree in
RA per day in the direction of the Galactic plane, then the sky patch
we assume can be considered representative of an early launch scenario
when the impact of Galactic foregrounds will be at their lowest.
Following~\cite{2011JCAP...08..001F}, we assume a three band
experimental configuration with channels at 150, 250 and 410 GHz
observing to depths of 5.25, 14.0 and 140 $\muK_{\rm CMB}-$arcmin
respectively, each with an angular resolution of $8'$. An advantage of
this selection of bands is to minimize the possible impact polarized
synchrotron, assumed to be negligible in this study, which may affect
ground-based CMB polarization experiments observing at 90
GHz~\cite{Stivoli_etal_2010}.

\begin{figure}
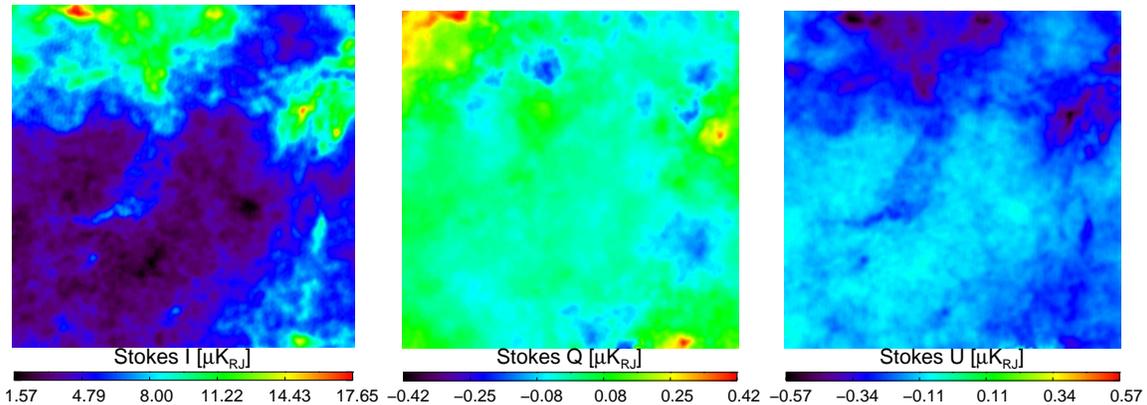

  \centering 
\includegraphics[width=1.95in]{\figdir{dust150_map0.eps}}
  \includegraphics[width=1.95in]{\figdir{dust150_map1.eps}}
  \includegraphics[width=1.95in]{\figdir{dust150_map2.eps}}
  \caption{Dust simulation (Stokes $I$, $Q$, and $U$) at 150 GHz,
  with a polarization fraction $p=0.036$, on a $13^\circ \times
  13^\circ$ patch centred on ($l$,$b$)=($250^\circ$,$-38^\circ$).}
  \label{fig:dustsims}
\end{figure}

Our Galactic polarized dust model is the same as the one first
described in~\cite{Stivoli_etal_2010}: dust intensity is given by the
model of~\cite{Schlegel_1998} extrapolated to 410 GHz. Then, to
simulate polarized emission, polarization angles are set on large
angular scales using the WMAP dust template~\cite{Page_etal_2007},
while on smaller scales, extra Gaussian power is added using the
prescription of~\cite{Giardino_2002}. The polarization fraction, $p$,
is assumed to be spatially constant, and we investigate three cases of
3.6, 5, and 10\%, intended to bracket the average high Galactic
latitude dust polarization detected in the WMAP W
band~\cite{2007ApJ...665..355K}, and possible higher dust polarization
fractions observed by ARCHEOPS at 353
GHz~\cite{2004A&A...424..571B}. The dust is scaled from the 410 GHz
band to the lower frequency bands assuming a greybody frequency
scaling
\begin{eqnarray}
A_{\rm dust}\propto\frac{{\nu}^{\beta+1}}{\exp \frac {h\nu}{kT}-1} ,
\label{eqn:dust}
\end{eqnarray}
with $T=18$K and $\beta=1.65$, with the dust temperature and
spectral index both assumed to be uniform across the patch. The
resulting dust polarization simulation at 150 GHz is shown in
Figure~\ref{fig:dustsims}.

For our CMB simulations, we produced two sets of 100
realizations--lensed and unlensed-- with $0.76'$ pixel size, assuming
the WMAP 7-year best-fit cosmological parameter
values~\citep{Komatsu_WMAP7}, and with our fiducial CMB
polarization power spectra calculated using {\sc CAMB}~\citep{camb}.

\begin{figure}
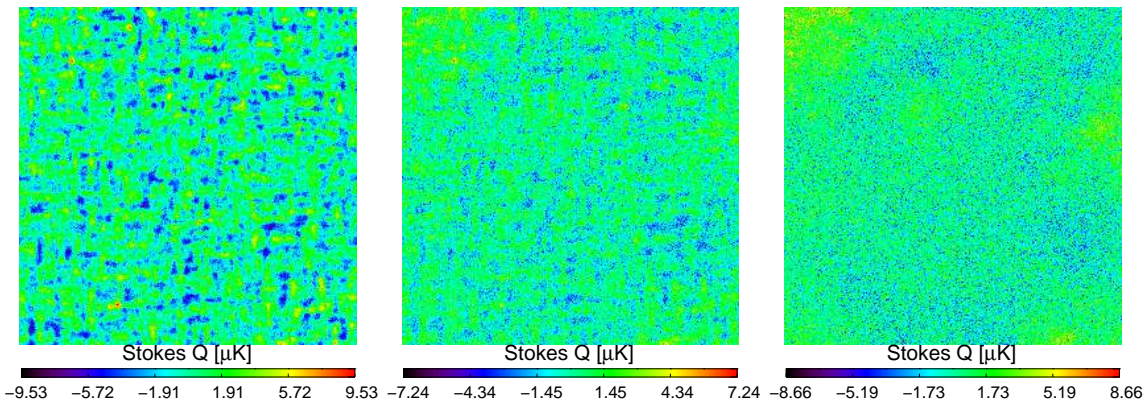

\centering 
\includegraphics[width=1.95in]{\figdir{sim150_map1.eps}}
\includegraphics[width=1.95in]{\figdir{sim250_map1.eps}}
\includegraphics[width=1.95in]{\figdir{sim410_map1.eps}}
\caption{From left to right, we show our CMB+dust+noise simulations at
150, 250 and 410 GHz for a dust polarization fraction $p=0.1$. The 150
GHz channel is CMB dominated, while the 410 GHz channel is dust
dominated. The 250 GHz channel provides information about the dust
spectral index.}
 \label{fig:cmbsims}
\end{figure}

The first set of maps--the lensed CMB realizations--were obtained
starting from the unlensed fiducial power spectrum $C_{\ell}^{XY}$,
from which Gaussian realizations of the CMB polarization were
generated, which were then lensed by remapping the pixels by the
deflection field.
The deflection field is in turn is derived
from a Gaussian realization of the projected potential power spectrum
$C_{\ell}^{\phi\phi}$; we neglect the effect of the integrated
Sachs-Wolfe effect induced correlation $C_{\ell}^{X\phi}$. We have
chosen our pixel size to be small compared to the RMS of the
deflection angles ($\sim2'$) so that errors due to interpolation back
onto the regular grid after remapping are small. We have checked that
the $E$ and $B$-mode power spectra of these simulated lensed maps
reproduces the lensed power spectra obtained from CAMB to within a few
percent accuracy for the $E$-mode spectrum and to within five percent
accuracy for the $B$-mode spectrum. While this is less accurate than
the all-sky lensing simulations now performed by several groups using
various interpolation schemes~\cite{Lewis_LensedCMB_05}
\cite{2009A&A...508...53B} \cite{2010ApJS..191...32L}
\cite{2012arXiv1205.0474B}, we believe that our flat-sky simulations
are sufficiently accurate for our dust foreground study.

The second set of maps--the unlensed CMB realizations--were obtained
from Gaussian realizations of the lensed fiducial power spectrum, $\tilde
C_{\ell}^{XY}$. These maps have the same power spectrum as the lensed
CMB realizations, but have none of the lensing-induced
non-Gaussianity.  Since the power spectrum of the convergence
reconstructed on unlensed CMB maps is same as the lensing noise power
spectrum predicted analytically from~\eqn{eq:nps}, these maps have
been used for checking the accuracy and implementation details of
the convergence and power spectrum estimators, as well as the testing
effect of mask apodization.

Finally the CMB maps are scaled to antenna temperature units in the
three bands at 150, 250 and 410 GHz, smoothed with an $8'$ beam, and
uncorrelated Gaussian white noise is added to each pixel. The Stokes
$Q$ parameter of an example simulation is shown in
Figure~\ref{fig:cmbsims}.

\section{Results}

This section describes our results in which we calculate the level of
lensing bias that is expected from our dust polarization model.
Finding a dust bias to be present, we apply the parametric
multi-frequency foreground cleaning technique~\cite{Stompor_etal_2009}
described in Section~\ref{sec:fgclean}, and establish a requirement
for the accuracy with which the dust spectral index must be
constrained in order to guarantee dust cleaning.

\subsection{Foreground-free case}

\begin{figure}
\centering
\includegraphics[width=5in]{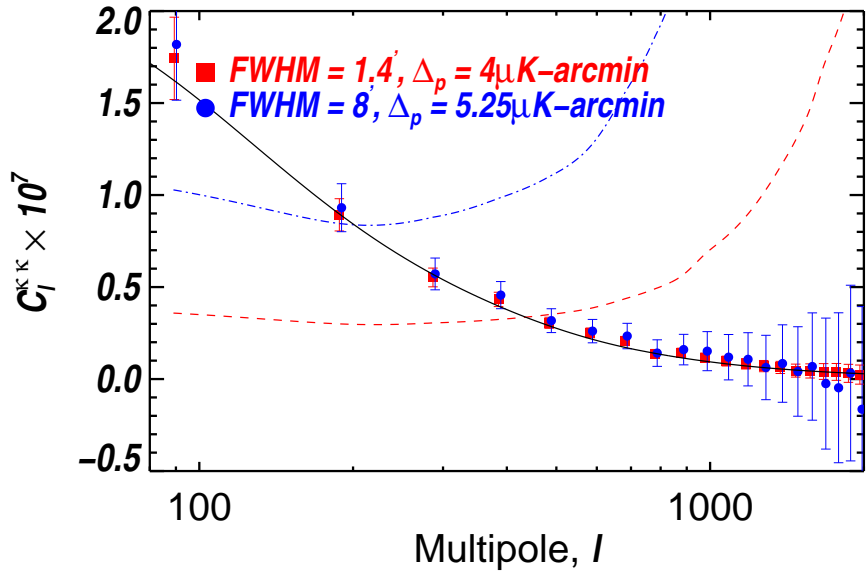} 
\caption{Validation of the $EB$ quadratic estimator. The solid line is
the power spectrum of the input convergence map
$C_{\ell}^{\kappa\kappa}$, and the error bars show the estimated
convergence power spectra averaged over 100 simulations. The dash
curves are the lensing noise spectrum, $\Nkap_{\ell}$, for the two
sets of survey parameters assumed.
}\label{fig:validate_cmblens}
\end{figure}

Before reporting the effect of the dust, we first demonstrate our
reconstruction of the convergence from the $EB$ quadratic estimator
under the most idealized foreground-free case. We used the CMB modes
in the range $l_{\rm min}<\ell<l_{\rm max}$, where the minimum
multipole is chosen to be twice the Nyquist mode, $k_{\rm nyq} =
\pi/\Delta \theta$, where $\Delta \theta$ is the angular size of the
patch in radians, while the maximum multipole is determined by the
noise level and beam size of the experiment. For our case $l_{\rm min}
=28$, and $l_{\rm max} = 3000$. As we will show later, the choice of
$l_{\rm min}$ becomes important when investigating the effect of
foregrounds, while varying $l_{\rm max}$ does not significantly change
our results.  Figure~\ref{fig:validate_cmblens} shows the convergence
power spectrum for our simulated 150 GHz channel with 5.25
$\muK-$arcmin sensitivity and 8$'$ angular resolution, as well as for
a survey with 4$\muK-$arcmin sensitivity and 1.4$'$ angular resolution
similar to the planned `ACTPol Deep' survey
of~\cite{2010SPIE.7741E..51N}. The power spectrum estimates shown are
the average over 100 simulations of $\hat C^{\kappa\kappa}_{\ell} -
\Nkap_{\ell}$, while the corresponding error bar is given by
\begin{equation}
\Delta C^{\kappa\kappa}_{\ell} =
       \frac{\hat C_{\ell}^{\kappa\kappa}+\Nkap_{\ell}}
       {\sqrt{\ell\Delta\ell f_{\rm sky}}},
\end{equation}
where we have applied a binning scheme $\Delta l=98$ (thirty bins
between $\ell=40$ and $\ell=3000$). The power spectrum estimates shown
in Figure~\ref{fig:validate_cmblens} represent an end to end
validation of our CMB simulations, $EB$ quadratic estimator, and power
spectrum estimation pipeline.

\subsection{Dust polarization bias at 150 GHz}

\begin{figure}
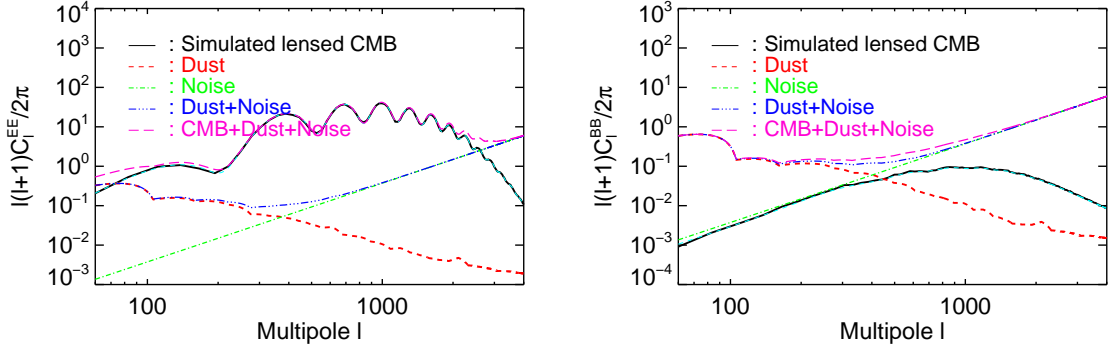

\centering
\includegraphics[width=3in]{\figdir{Cl_EE_cmb_dust01_noise5pt25_150GHz.eps}}
\includegraphics[width=3in]{\figdir{Cl_BB_cmb_dust01_noise5pt25_150GHz.eps}}
\caption{The lensed CMB $E$-mode (left panel) and $B$-mode (right
panel) power spectrum is compared to the power spectrum of the dust
($p=0.1$) and instrumental noise at 150 GHz.  For comparison the
fiducial $E$ and $B$-mode power spectra is also shown (cyan).}
 \label{fig:dust_spec}
\end{figure}

To first assess the size of the dust contamination on the patch we are
considering, we estimated the power spectrum of the simulated dust at
150 GHz and compared it to $E$ and $B$-mode signal and noise power
spectra, as shown in Figure~\ref{fig:dust_spec} for the example of
$p=0.1$. For our dust model and choice of patch, the $E$ and $B$-mode
power spectra of the dust approximately follow a powerlaw given by
$C_{\ell}^{\rm dust}=(A\times p)^2\ell^{\beta}$, where $p$ is the
polarization fraction, $A\simeq120\mu$K and $\beta\simeq-3.5$.  Our
previous study~\citep{2011JCAP...08..001F} has shown that polarized
dust at this level of power must be modeled and subtracted in order to
derive unbiased estimated of the inflationary $B$-mode spectrum, a
cosmological signal which is accessible in the $\ell<200$ range of the
$B$-mode power spectrum. The main question we seek to address in this
study is whether this level of anisotropy power of foreground
contamination is large enough to also bias the estimates of the
lensing signal.
\begin{figure}[H]
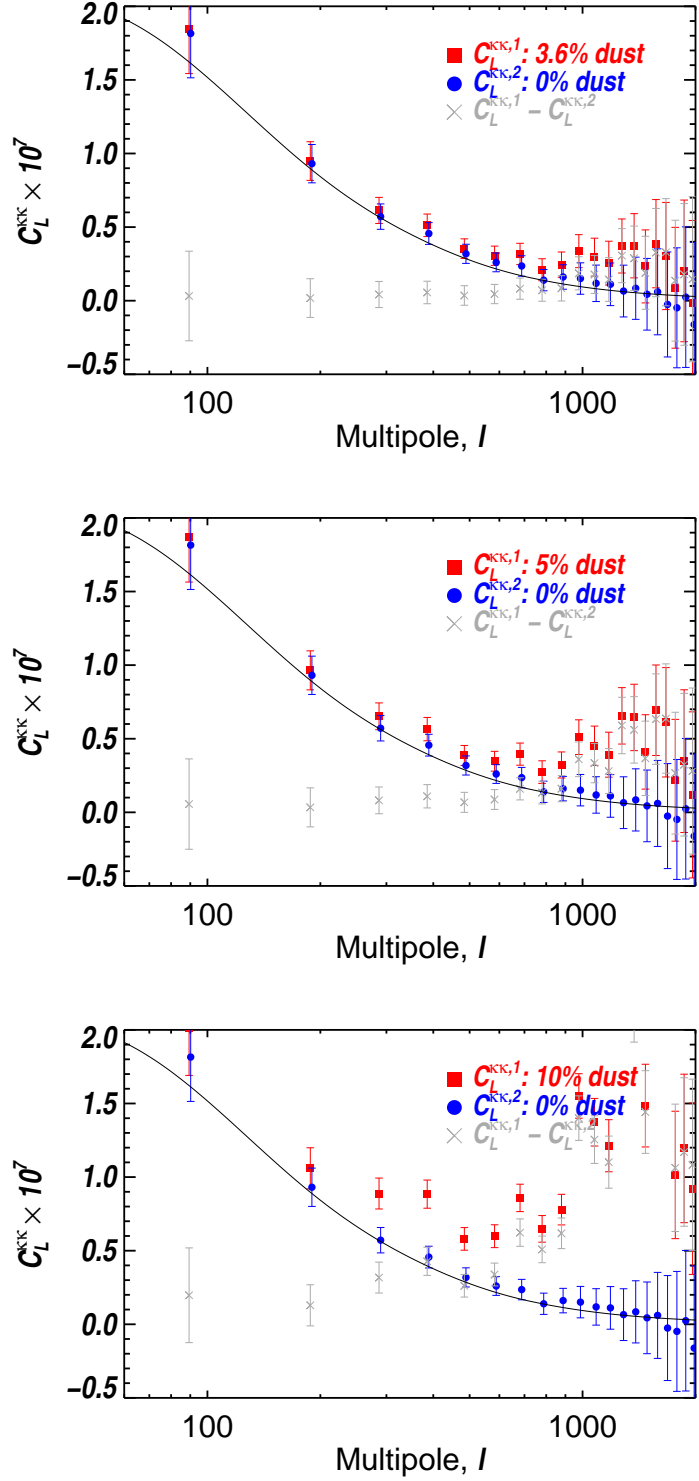

\centering
\includegraphics[width=4in]{\figdir{clkk_lmin20_dust0036_fwhm8_sigpix52_ignored.eps}} 
\includegraphics[width=4in]{\figdir{clkk_lmin20_dust005_fwhm8_sigpix52_ignored.eps}} 
\includegraphics[width=4in]{\figdir{clkk_lmin20_dust01_fwhm8_sigpix52_ignored.eps}}
\caption{Effect of polarized dust at 150 GHz for surveys with
sensitivity 
$\Delta_{\rm p} = 5.25\mu{}K-$arcmin.  The upper, middle and lower  
 panels show cases with dust polarization fractions of
 $p=0.036$, 0.05 and 0.1 respectively. 
We find that for our dust model and choice of patch, 
diffuse polarized dust is expected to be a
significant source of bias to $C_{\ell}^{\kappa\kappa}$ for dust
polarization fractions of a few percent.}
 \label{fig:dust_band1}
\end{figure}

We have calculated the power spectrum of the convergence field
reconstructed with the $EB$ quadratic estimator using the dust
contaminated 150 GHz channel, and show our results in
Figure~\ref{fig:dust_band1} for three different polarization fractions
of  $p= [0.036, 0.05, 0.1]$.  We find that if the dust contamination is
ignored during the lensing estimation, then a `dust noise bias'
dominates over the lensing power spectrum estimates for both the
$p=0.05$ and $p=0.1$ case,  with only a slight excess for the $p=0.036$ case.

Although our demonstration of the dust bias will be dependent on the
choice of patch we have assumed, and on the details of our polarized
dust model and its power spectrum, we nonetheless conclude that
diffuse polarized dust may in principle be a source of bias for future
sub-orbital CMB surveys aiming at lensing estimates using the $EB$
estimator, and that methods for foreground debiasing must therefore be
developed.

{For our dust model and choice of patch we can now roughly estimate
that the requirement on the `dust suppression factor' (the factor by
which the polarized dust must be reduced at 150 GHz) is approximately
$(p_{\rm crit}/p)$, where we have estimated that the `critical
polarization fraction' $p_{\rm crit}$, below which the effect of dust
contamination becomes small compared to the noise level in the
convergence power spectrum estimates, is approximately $p_{\rm
crit}=0.01$. This requirement on the dust suppression factor, $(p_{\rm
crit}/p)$, will in turn set the requirement with which the dust spectral
index must be estimated, which we will calculate in the next section.}

\subsection{Debiasing the effect of dust via component separation}

Having established that our polarized dust model is leading to a bias
in the lensing power spectrum, our main aims are 1) To test the
multi-frequency foreground cleaning technique described in
Section~\ref{sec:fgclean}, in which the dust spectral index is first
estimated from the data using the profile likelihood
Eq.~(\ref{eqn:profile}), before applying a linear least-squares
component separation of the CMB and dust using Eq.~(\ref{eqn:gls}) and
propagation of the noise covariance using Eq.~(\ref{eqn:noisecorr}),
and 2) To put requirements on the accuracy with which the dust
spectral index must be estimated in order to guarantee dust cleaning.

For the homogeneous noise case that we have simulated, the final least
squares combination of the data is a linear combination of the channel
maps. However, the profile likelihood is a non-linear method involving
quadratic combinations of the data, and so we have therefore
numerically investigated its accuracy in determining the dust spectral
index by carrying out 100 spectral index estimation and component
separation simulations, each with a different CMB and noise
realization.

Table~\ref{tab:beta_d} summarizes our results for our dust spectral
index estimation simulations, for three cases
$p=[0.036,0.05,0.1]$. First, we find that the greater the dust signal,
the more accurately the dust spectral index can be estimated from the
data. This is qualitatively consistent with the analysis of the
profile likelihood of~\cite{2011PhRvD..84f3005E} in which increased
dust contrast is shown to improve the accuracy of the dust spectral
index estimation. We have also found that the profile likelihood
returns an unbiased estimate for the dust spectral $\beta$ when
considering the average value obtained over the ensemble of CMB+noise
realizations that we have analysed, but we also find that the profile
likelihood width for any single noise underestimates the dust spectral
index uncertainty. This breakdown of the profile likelihood appears to
be related to the relatively low contrast regime of the polarization
data we have simulated: If the dust intensity data is assumed to have
the same frequency scaling as the polarization data, then the dust
spectral index is first estimated with much higher accuracy, and the
profile likelihood width and Monte Carlo average of the spectral index
estimates agree well. This implies that any application of the
profile likelihood to data should be accompanied by Monte Carlo
simulations, given a Galactic foreground model, in order to more
accurately quantify the spectral index uncertainty.

For each of the 100 spectral index estimates, we calculate the
$Z$-matrix defined in Eq.~(\ref{eqn:zmat}); We will focus on the
$Z_{\rm CMB, Dust}$ matrix element, since this will quantify the
fraction of dust that remains unsubtracted from the CMB component
after the least squares component separation. Table~\ref{tab:zmat}
gives the mean and RMS of $Z$-matrices for the three
polarization fraction cases $p=[0.036,0.05,0.1]$. The basic conclusion
is that while in the case of $p=0.1$ the dust spectral index is always
estimated accurately enough to guarantee dust cleaning ($\left<Z_{\rm CMB,
Dust}\right>=0.00\pm0.10$), in the case of $p=0.036$ the dust is no longer
bright enough to allow sufficiently accurate spectral index
estimation, and for many simulations, there is an \emph{amplification}
of the dust contamination ($|Z_{\rm CMB, Dust}|> 1$).

\begin{table}
\begin{center}
\begin{tabular}{cccc}
($\beta_{\rm input}=1.65$)  & & \\
\hline
\hline
 $p$ & $\beta$ & $\Delta \beta$ \\
\hline 
 0.036 & 1.69   &   0.43 (0.04)  \\
 0.05 & 1.67   & 0.22 (0.03) \\
 0.1 & 1.66  &  0.06 (0.02) \\
\hline
\end{tabular}
\end{center}
\caption{Mean and RMS of the dust spectral index estimate $\beta$,
estimated from 100 component separation simulations varying CMB and
noise realizations, for the three dust polarization fraction cases
assumed;  The values in parentheses are the mean value of the profile
likelihood width. We find that spectral index uncertainty for any
single noise realisation (shown in parentheses) is underestimated by
the profile likelihood in the dust contrast regime of our polarization
simulations.}\label{tab:beta_d}
\end{table}

\begin{figure}[H]
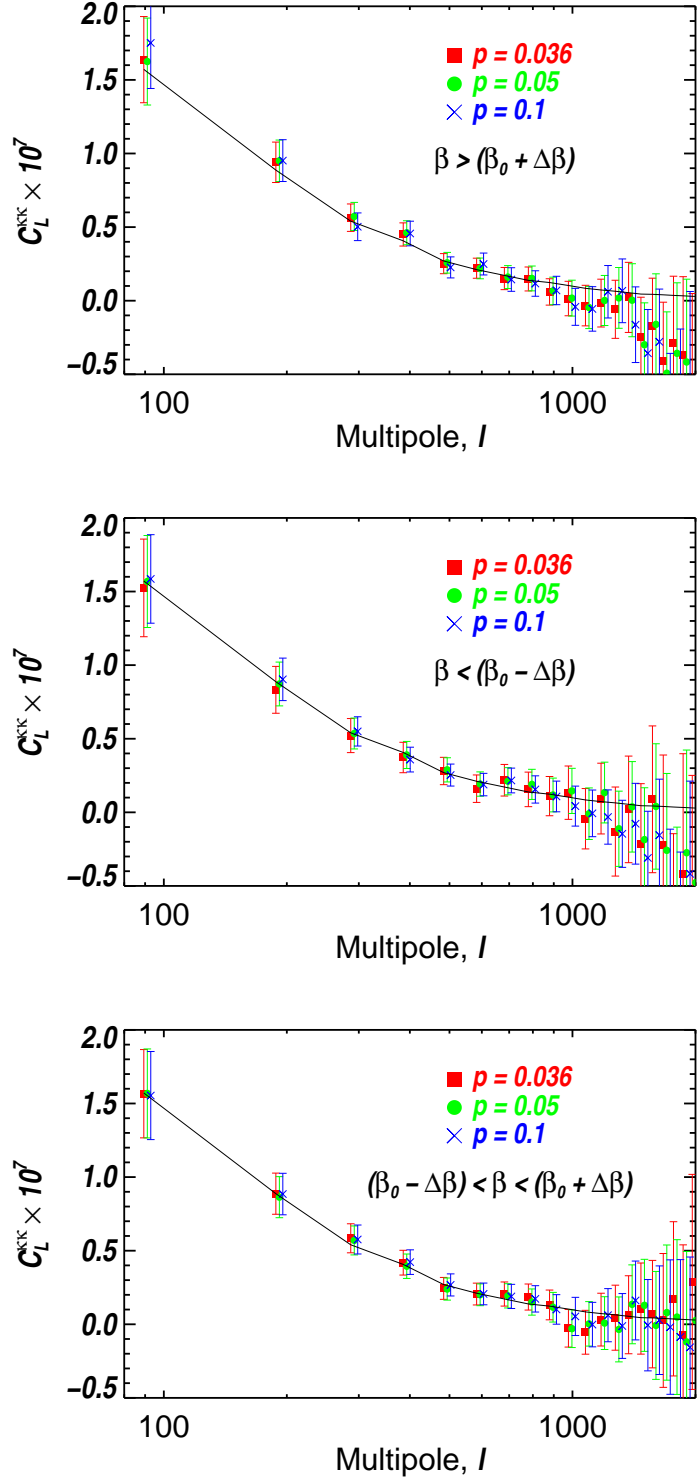

\centering
\includegraphics[width=4in]{\figdir{Clkk_beta_up_fgclean_dust_all_cases.eps}}
\includegraphics[width=4in]{\figdir{Clkk_beta_down_fgclean_dust_all_cases.eps}}
\includegraphics[width=4in]{\figdir{Clkk_beta_middle_fgclean_dust_all_cases.eps}}
\caption{Convergence power spectrum of the estimated CMB
component. The upper and middle panels shows the average
convergence power spectrum for realisations with spectral index
$\beta>[1.81,1.76,1.71]$ and $\beta<[1.49,1.54,1.59]$, respectively,
for the three dust polarization fraction cases of
$p=[0.036,0.05,0.1]$. The lower panel shows the average convergence
power spectrum for those CMB components whose spectral index meets the
accuracy requirement given in Eq.~(\ref{eq:beta_criterion}).}
\label{fig:residuals}
\end{figure}

It therefore becomes important to be able to put a requirement on the
accuracy, $\Delta\beta$, with which the dust spectral index must be
estimated in order to guarantee foreground cleaning. This requirement
can then either guide the application of our parametric component
separation technique, or inform the usage of possible external
spectral index prior information. For our case study we suggest the
criterion
\begin{equation}
\Delta \beta < \left(\frac{p_{\rm crit}}{p}\right)\times\left.\frac{\partial\beta}{\partial Z}\right|_{\beta_0},
\label{eq:beta_criterion}
\end{equation}
where we conservatively assume $p_{\rm crit}=0.01$; More generally,
the factor $(p_{\rm crit}/p)$ can be thought of as the desired dust
suppression factor for a given patch of sky. Qualitatively, there is
therefore a requirement for greater spectral index precision for
larger values of $p$. Quantitatively, we have numerically calculated
$\left.\frac{\partial\beta}{\partial Z}\right|_{\beta_0}=0.56$ from
the function $Z(\beta)$, and so for the three cases
$p=[0.036,0.05,0.1]$ we require $\Delta\beta<[0.16,0.11,0.06]$
respectively. We have checked that enforcement of this spectral index
accuracy requirement leads to satisfactory dust
cleaning. Figure~\ref{fig:residuals} shows the average convergence
power spectrum of the estimated CMB component for those realizations
whose spectral index estimate meet this accuracy requirement (for
instance those with $\beta=1.65\pm0.06$ for the $p=0.1$ case), for which we find
satisfactorily debiased convergence power spectrum estimates after
component separation. Conversely, the realizations with estimated dust
spectral indices falling outside the required range show a biased
convergence power spectrum after component separation.

\begin{table}
\begin{center}
\begin{tabular}{cccc}
$p = 0.036$ \hfill\hfill & & \\
\hline
\hline
Input:\hfill\hfill & CMB & Dust  \\ 
\hline
Output:\hfill\hfill & & & \\
CMB & $1$   & $-0.06\pm 0.79$ \\
Dust & $0$  & $1.04\pm 0.41$ \\
\hline
$p=0.05$ \hfill\hfill & & \\
\hline
\hline
Input:\hfill\hfill & CMB & Dust  \\ 
\hline
Output:\hfill\hfill & & & \\
CMB & $1$   & $-0.01 \pm 0.41$ \\
Dust & $0$  & $1.01 \pm 0.21$ \\
\hline
$p=0.1$ \hfill\hfill & & \\
\hline
\hline
Input:\hfill\hfill & CMB & Dust  \\ 
\hline
Output:\hfill\hfill & & & \\
CMB & $1$   & $0.00 \pm 0.10$ \\
Dust & $0$  & $1.00 \pm 0.05$ \\
\hline
\end{tabular}
\end{center}
\caption{$Z$-matrices, Eq.~(\ref{eqn:zmat}), averaged over 100
  component separation simulations varying CMB and noise realizations,
  for three different dust polarization cases.  The $Z_{\rm CMB,
  Dust}$ matrix element quantifies the fraction of dust at 150 GHz
  that is mixed into the CMB after component separation, and so can be
  thought of as a `dust suppression factor'. We find that in the
  $p=0.1$ case, the dust spectral index is estimated with enough
  accuracy to guarantee dust cleaning ($|\left<Z_{\rm CMB,Dust}\right>|\ll 1$),
  while for the $p=0.36$ case dust cleaning is not guaranteed.}
\label{tab:zmat}
\end{table}

\subsection{Discussion}

\begin{figure}[]
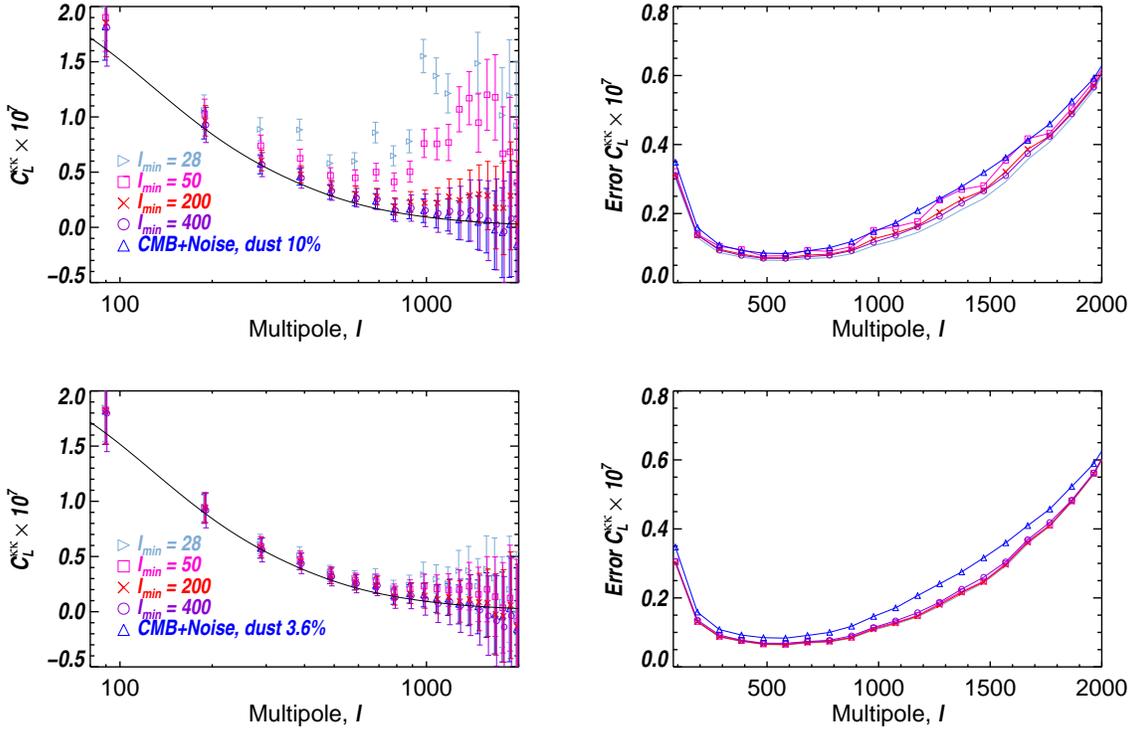

\centering
\includegraphics[width=3in]{\figdir{clkk_lmin_change_dust01_fwhm8_sigpix52_ignored.eps}} 
\includegraphics[width=3in]{\figdir{error_clkk_lmin_change_dust01_fwhm8_sigpix52_ignored.eps}}
\includegraphics[width=3in]{\figdir{clkk_lmin_change_dust0036_fwhm8_sigpix52_ignored.eps}} 
\includegraphics[width=3in]{\figdir{error_clkk_lmin_change_dust0036_fwhm8_sigpix52_ignored.eps}}
\caption{The trade-off between dust bias and the noise level when
  varying the value of $l_{\rm min}$, for $p=0.1$ (upper panels) and
  $p=0.036$ (for lower panels).  The left column shows the convergence power spectrum
  and the right column shows the corresponding power spectrum error bars.
  For the cases we studied, the $l_{\rm min}$ required to yield unbiased
  convergence power are approximately $[100,200,400]$ for $p=[0.036,0.05,0.1]$,
  respectively. Low $l$ filtering provides a useful robustness check
  for the effects of diffuse foreground biases.}
 \label{fig:clkk_lmin}
\end{figure}

In the previous section we have shown that the lensing signal may well
be subject to biases from diffuse polarized dust, for which we have
investigated a possible mitigation strategy in the form of a
parametric component separation of CMB and dust. The generality of
this result and foreground mitigation strategy is worth questioning:
are there other options for the manner in which we may trade dust bias
for increased variance?

Firstly, \reffig{fig:dust_spec} shows that the dust contamination has
a `red' anisotropy power spectrum, and this suggests a possible
strategy for mitigating the dust bias. As long as we have information
about the power spectrum of dust, then filtering the low-multipole
modes can be used to reduce the
bias~\cite{Hu:2007njp,Engelen:2012spt}, perhaps at an acceptable cost
to the variance. We have demonstrated this technique by varying
$l_{\rm min}$, and found that appropriate tuning of this parameter can
indeed reduce the dust bias effect. Specifically we found that the
value of $l_{\rm min}$ that results in unbiased estimates of the
convergence power spectrum depends on the polarization fraction of the
foreground: the greater the foreground level, the more aggressive the
required low-multipole filtering. For our dust model and choice of
patch, the approximate required filtering scale is given by $l_{\rm
  min} \sim 100\times (p/0.036)$. This is illustrated in
\reffig{fig:clkk_lmin} which shows the convergence power spectrum and
lensing noise level as a function of $l_{\rm min}$ cut, for the cases
of dust polarization fraction of $p=0.1$ and $p=0.036$. We judged that
the values of $l_{\rm min}$ required to yield unbiased convergence
power spectra, which is obtained by demanding the $\chi^2$ with
respect to the input convergence power spectrum is less than the
$\chi^2$ obtained for the $1\%$ dust polarization case, are
$[100,200,400]$ for $p=[0.036,0.05,0.1]$, respectively. For dust
polarization fractions less than $0.05$, we found that the loss in the
signal due to the $l_{\rm min}$ cut is small enough to yield error
bars close to the foreground-free case.

\begin{figure}[]
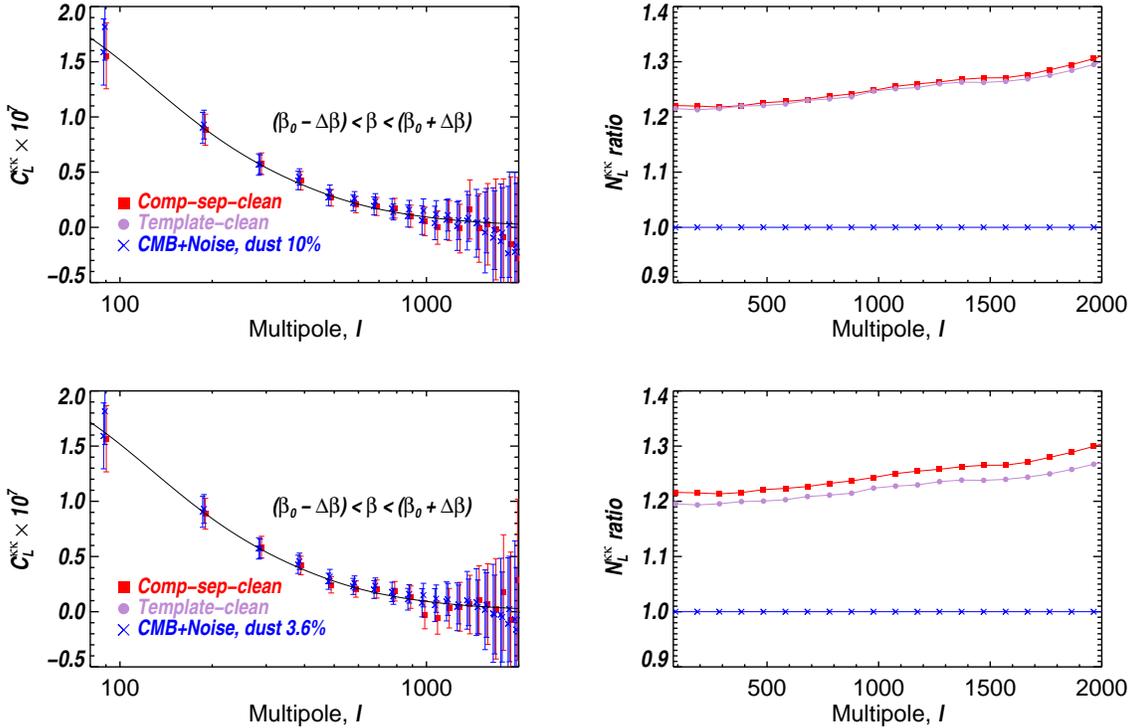

\centering
\includegraphics[width=3in]{\figdir{Clkk_beta_middledust01_fgclean_dust_all_cases.eps}}
\includegraphics[width=3in]{\figdir{Nlkk_beta_middledust01_fgclean_dust_all_cases.eps}}
\includegraphics[width=3in]{\figdir{Clkk_beta_middledust0036_fgclean_dust_all_cases.eps}}
\includegraphics[width=3in]{\figdir{Nlkk_beta_middledust0036_fgclean_dust_all_cases.eps}}
\caption{Comparison of parametric component separation and
   template-based foreground cleaning methods. The left and right
   columns show the convergence power spectra and lensing noise ratios
   (relative to the foreground-free case) respectively, while the
   upper and lower rows are for the $p=0.1$ and $p=0.036$ cases
   respectively.  Template cleaning may provide a useful alternative
   to and consistency check of parametric component separation, where
   the final noise level of the cleaned CMB estimate is fairly
   insensitive to the level of foreground contamination.
}
 \label{fig:clkk_tclean}
\end{figure}

Secondly, in the light of possible component separation biases
discussed in the last section, it is worth considering whether
template-based
methods~\citep{2007ApJS..170..335P,2009MNRAS.397.1355E,2011ApJ...737...78K}
may be useful for foreground cleaning in this context. Note that
template-based methods typically have fewer free parameters to
be estimated than the component separation method that we considered, in
which both dust amplitude and frequency scaling parameter are
estimated pixel by pixel; this is likely to be at the heart of the
difficulty of applying a full component separation.  We therefore
tested an approximate template-based cleaning method in which the
dust-dominated 410 GHz channel is used as a polarized dust template to
suppress the foreground contamination in the 150 GHz channel. The dust
amplitude coefficient, $\alpha_d$, is estimated by maximising the
likelihood
\begin{equation}
-2 \ln \mathcal{L} = \sum_p
\frac{(Q_{150}- \alpha_d\times Q_{410})^2}{\sigma_{Q, 150}^2}+
\frac{(U_{150}- \alpha_d\times U_{410})^2}{\sigma_{U, 150}^2},
\end{equation}
where the pixel size has first been degraded to $6.1'$. Once the template
coefficient has been estimated then the full resolution maps are
appropriately combined, and the noise is propagated using
\begin{eqnarray}
\hspace{0cm} [\sigma_Q^2,
  \sigma_U^2] &=& \frac{[\sigma_Q^2,\sigma_U^2]^{150}
  +\alpha_d^2[\sigma_Q^2,\sigma_U^2]^{410}}{(1-\alpha_d)^2}.
\end{eqnarray}
\reffig{fig:clkk_tclean} shows our results from our template cleaning
and compares them to the performance of the component separation
method. Our basic finding is that template cleaning may provide a
useful alternative to and consistency check of parametric component
separation, and one that is robust in the sense that the final noise
level of the cleaned CMB estimate is fairly insensitive to the level
of foreground contamination.

\section{Conclusions}

Several ongoing and planned CMB polarization experiments are aiming to
measure and characterise the lensing of the cosmic microwave
background, in order to improve constraints on the parameters of the
cosmological model. Within this context we have made the first
specific study of the possible effect of diffuse polarized dust
emission on the accuracy of the reconstruction of the lensing
convergence signal. Our particular focus has been on performing a case
study of a three channel balloon-borne CMB experiment covering the
frequency range 150--410 GHz. Our numerical investigation is based on
a dust polarization simulation and a flat-sky implementation of the Hu
and Okamoto quadratic estimator. We found that for the sky patch under
consideration, which is near to the region of sky that will be
targeted by the EBEX  experiment, and for plausible dust
polarization fractions in the range 3.6--10\%, the anisotropy of the
diffuse dust polarization will be large enough at 150 GHz to bias the
reconstruction of the convergence. Thus a multi-frequency experimental
approach is imperative, and appropriate analysis methods must be
developed for debiasing the effect of polarized dust.

In order to mitigate the effect of the dust and to debias the
convergence power spectrum, we demonstrated that a multi-frequency
component separation technique in which the dust spectral index is
first estimated from the data using a profile likelihood technique,
before applying a least squares component separation. We found
evidence for a dust contrast regime in which the accuracy of profile
likelihood breaks down, both underestimating the spectral index
uncertainty as well as providing spectral index estimates that are
insufficiently accurate to guarantee dust cleaning. This highlights
the possible need for external constraints on the frequency scaling of
the polarized dust which can then be used as a prior, to stabilize the
dust spectral index estimates to the accuracy required for sufficient
dust cleaning. We proposed a criterion, Eq.~(\ref{eq:beta_criterion}),
which sets a requirement for the accuracy with which the spectral
index of the foregrounds must be estimated in order to guarantee a
given dust suppression factor. We then demonstrated that satisfactory
dust cleaning was achieved for the cases in which the estimated
spectral index met this requirement. Given these concerns, we
showed that removing the lower-multipole foreground-contaminated CMB
modes from the lensing reconstruction, as well as using the 410 GHz
channel as a dust template provide two further methods for diffuse
foreground mitigation. We expect, though, that detailed parametric
modeling of the frequency scaling of foregrounds will be important for
removing possible foreground-coupled systematic effects that may
affect the forthcoming half-wave plate polarimeters designed to
measure $B$-mode polarization~\citep{2012ApJ...747...97B}.

\acknowledgments APSY acknowledges support from NASA grant number
NNX08AG40G. CB acknowledges partial support from the PD51 INFN
grant. We acknowledge the use of the {\sc
HEALPIX}~\cite{2005ApJ...622..759G} and {\sc CAMB}~\cite{camb}
packages.

\bibliography{cmblens}

\end{document}